\documentclass{llncs}

\usepackage{code}
\usepackage{amsmath, amssymb, theorem}
\usepackage{stmaryrd}

\theorembodyfont{\rmfamily\mdseries\upshape}

\newcommand{\mysection}[1]{\vspace*{-4mm}\section{#1}\vspace*{-2mm}}
\newcommand{\mysubsection}[1]{\vspace*{-3mm}\subsection{#1}\vspace*{-1mm}}

        {\hfill $\Diamond$\end{ex}}

\newcommand{\ignore}[1]{}

\renewcommand{\alpha}{a}
\renewcommand{\beta}{b}
\renewcommand{\gamma}{c}
\renewcommand{\tau}{t}

\newcommand{\fset}[1]{\left\{#1\right\}}

\newcommand{\pjs}[1]{\marginpar{\sc pjs}{\bf #1}}
\newcommand{\ms}[1]{\marginpar{\sc ms}{\bf #1}}

\newcommand{\comment}[1]{}

\renewcommand{\implies}[0]{\supset}
\newcommand{\arrow}[0]{\rightarrow}

\newcommand{\gd}[0]{~\rule{0.5mm}{2.5mm}~}

\newcommand{\bi}{\begin{array}[t]{@{}l@{}}}
\newcommand{\ei}{\end{array}}
\newcommand{\ba}{\begin{array}}
\newcommand{\ea}{\end{array}}
\newcommand{\bda}{\[\ba}
\newcommand{\eda}{\ea\]}
\newcommand{\bp}{\begin{quote}\tt\begin{tabbing}}
\newcommand{\ep}{\end{tabbing}\end{quote}}

\newcommand{\alphavec}{\bar{\alpha}}

\newcommand{\tenv}{\Gamma}

\newcommand{\tlabel}[1]{\mbox{(#1)}}
\newcommand{\fig}[3]
        {\begin{figure*}[t]#3\
        \caption{\label{#1}#2}\ \hrulefill\ \end{figure*}}

\newcommand{\myirule}[2]{{\renewcommand{\arraystretch}{1.2}\ba{c} #1
                      \\ \hline #2 \ea}}

\newcommand{\tv}{\mbox{\it fv}}

\newcommand{\turns}{\, \vdash \,}

\newcommand{\trl}{\, \vdash_{\scriptsize \id{W}} \,}

\newcommand{\id}[1]{\mbox{\it #1}}

\newcommand{\sgap}{\quad}

\newcommand{\mathem}{\sf}
\newcommand{\IN}{\mbox{\mathem in}}

\newcommand{\CASE}{\mbox{\mathem case}}

\newcommand{\OF}{\mbox{\mathem of}}

\newcommand{\proofsin}[1]{#1}

\title{Type Inference for Guarded Recursive Data Types}

\author{Peter J. Stuckey\inst{1} and Martin Sulzmann\inst{2}}

\institute{
       NICTA Victoria Laboratory \\
       Department of Computer Science and Software Engineering\\
       The University of Melbourne, Vic.\ 3010, Australia\\
       \email{pjs@cs.mu.oz.au}
\and
        School of Computing, National University of Singapore \\
        S16 Level 5, 3 Science Drive 2, Singapore 117543 \\
        \email{sulzmann@comp.nus.edu.sg}}

\begin{document}

\maketitle


\bibliographystyle{alpha}

\begin{abstract}
We consider type inference for guarded recursive data types (GRDTs) -- 
a recent generalization of algebraic data types.
We reduce type inference for GRDTs to
unification under a mixed prefix. Thus, we obtain efficient type inference.
Inference is incomplete because 
the set of type constraints allowed to appear in the type system is only a subset of those 
type constraints generated by type inference.
Hence, inference only succeeds if the program is sufficiently type annotated.
We present refined procedures to infer types incrementally and to assist the user in
identifying which pieces of type information are missing.
Additionally, we introduce procedures to test if a type
is not principal and to find a principal type if one exists.
\end{abstract}


\mysection{Introduction}

Guarded recursive data types (GRDTs)~\cite{604150} were introduced
by Xi, Chen and Chen as generalization of algebraic data types.
The novelty of GRDTs is that we may include type equality assumptions to refine types on a per
constructor basis. Thus, we can type more programs.

\vspace{-1mm}
\begin{example} \label{ex:eval}
The following data type ensures type correct construction of a simple expression language.
Note that we make use of Haskell-style syntax in examples.
\begin{code}
data Exp  a = (a=Int) => IsZero  | (a=Int) => IsSucc (Exp Int)
            | forall b c. (a=(b,c)) => Pair (Exp b) (Exp c)
eval :: Exp a -> a
eval Zero = 0
eval (Succ e) = (eval e) + 1
eval (Pair x y) = (eval x, eval y)
\end{code}
In contrast to algebraic data types we may refine the type of a GRDT depending on the
particular constructor.
E.g., {\tt IsZero 0} has type $Exp~Int$ whereas {\tt Pair (IsSucc (IsZero 0) (IsZero 0)}
has type $Exp~(Int,Int)$.
At first look it may be surprising that {\tt eval} has type $\forall a.Exp~a \arrow a$.
Consider the first clause. We assume that {\tt Zero} has type $Exp~a$ where we temporarily make
use of $a=Int$. Hence, we can give {\tt 0} the type $a$.
Note that we make use of polymorphic recursion, see the third clause.
\end{example}
\vspace{-1mm}

The idea of GRDTs dates back to Zenger's index types~\cite{zenger:thesis}.
He introduces a variant of the Hindley/Milner system where types ranging
over indices can be refined for each constructor. 
Variants of GRDTs have been studied by a number of authors~\cite{first-class-phantom,gad,eq-types,simonet-pottier-hmg}
whereas inference has received so far little attention.
We are only aware of the work by
Simonet and Pottier~\cite{simonet-pottier-hmg} and 
Peyton-Jones, Washburn and Weirich~\cite{gad}.
Simonet and Pottier establish some sufficient conditions
under which type inference can be reduced to some tractable constraint solving.
Essentially, they demand that every polymorphic recursive function and every use
of a GRDT must be annotated.
A similar approach is pursued by Peyton-Jones, Washburn and Weirich~\cite{gad}.
In general, it is acceptable practice to 
demand some form of user-provided type information to support tractable type inference.
This may be in particular crucial in case of  polymorphic recursion~\cite{henglein}.
GRDT programs make often use of polymorphic recursion. Hence, there is no hope
to obtain complete type inference for GRDTs unless we provide type annotations.
However, we would like to minimize the amount of user-provided annotations
and if possible provide feedback to the programmer which pieces of information
are missing.

In this paper, we propose several novel strategies to support inference for GRDTs.
In summary, our contributions are:

\begin{enumerate}
 \item We introduce an efficient inference method for GRDTs based on a translation
       from program text to constraints where constraints are solved by unification under a mixed prefix.
       In case of (potentially polymorphic) recursive functions, we present a refined procedure
       which allows to supply inference with partial type information (Section~\ref{sec:inference}).
 \item We give a sufficient criteria under which constraint solving is guaranteed to succeed.
       Failure of the criteria may provide useful feedback to the programmer which type information
       must be user-provided (Section~\ref{sec:inf-criteria}).
 \item We introduce a method to construct solutions out of the individual results from 
       successful sub-branches (Section~\ref{sec:enumerate}).
 \item 
       We give an efficient but incomplete procedure to test if a type is {\em not} principal.
       Under some assumptions, we give a method to infer 
       a principal type (Section~\ref{sec:principal}) if one exists.       
\end{enumerate}

Proof sketches of our results can be found in the Appendix.


We assume the reader is familiar with the concepts
of substitutions, most general unifiers (m.g.mu.), unification under a mixed prefix,
skolemization and the basics of first-order logic.
We refer to~\cite{lassez-maher-marriott:unification,147067,Sho67} for more details.


\mysection{Guarded Recursive Data Types} \label{sec:grdts}

\newcommand{\REC}{\mbox{\mathem rec}}

\fig{f:type-system}{Typing Rules}{  
\bda{ccc}
\ba{cc}
\tlabel{Eq} &
 \myirule{C, \tenv \turns e : t \\ C \implies t=t'}
         {C, \tenv \turns e : t'}
\ea
 &
  \ba{cc}
    \tlabel{Var-x} &
    \myirule{(x:\forall\bar{a}.C' \Rightarrow t) \in \tenv \\
              C \implies [\overline{t/a}]C'}
          {C, \tenv \turns x:[\overline{t/a}]t}
  \ea
 &
\ba{cc}
  \tlabel{Rec} &
 \myirule{\tenv.f:t \turns e : t}{\tenv \turns \REC f\, \IN\ e: t}
\ea
\eda
\bda{cc}
   \ba{cc}
    \tlabel{Abs} &
    \myirule{C, \tenv.x:t_1 \turns e:t_2}
          {C,\tenv \turns \lambda x.e: t_1 \arrow t_2}
   \ea
  &
    \ba{cc}
    \tlabel{App} &
    \myirule{C, \tenv \turns e_{1}:t_{2} \rightarrow t ~~~
           C, \tenv \turns e_{2}:t_{2}}
          {C, \tenv \turns e_{1} ~e_{2}: t}
    \ea
\eda
\bda{cc}

\ba{cc}
\tlabel{Case} &
 \myirule{ C, \tenv \turns e : t_1 \\
           C, \tenv \turns p_i \arrow e_i : t_1 \arrow t_2 \sgap \mbox{for $i\in I$}}
         {C, \tenv \turns \CASE\ e~ \OF\ [p_i \arrow e_i]_{i\in I} : t_2}

\ea
 &
\ba{cc}
 \tlabel{Pat} &
 \myirule{p:t_1 \turns \forall \bar{b}.(D \gd \tenv_p) \\
          \tv(C,\tenv,t_2)\cap \bar{b} =\emptyset \\
          C\wedge D, \tenv \cup \tenv_p \turns e : t_2}
         {C, \tenv \turns p \arrow e : t_1 \arrow t_2} 
\ea
\eda
\bda{cc}
 \ba{cc}
   \tlabel{Annot} & 
   \myirule{C_2 \wedge C_1, \tenv \turns e : t \sgap \tv(C_2,\tenv)\cap\tv(C_1,t)=\emptyset}
           {C_2, \tenv \turns (e::(C_1 \Rightarrow t)) : t}

 \ea
  &
 \ba{cc}
   \tlabel{Pat-Var} &
  x:t \turns (True \gd \{x:t\})
\ea
\eda
\bda{cc}
 \tlabel{Pat-Pair} &
  \myirule{p_1:t_1 \turns \forall \overline{b_1}.(D_1 \gd \tenv_{p_1}) \sgap
           p_2:t_2 \turns \forall \overline{b_2}.(D_2 \gd \tenv_{p_2})}
          {(p_1,p_2):(t_1,t_2) \turns 
      \forall \overline{b_1,b_2}. (D_1 \wedge D_2 \gd \tenv_{p_1}\cup \tenv_{p_1})}
\\ \\
 \tlabel{Pat-K} &
 \myirule{K:\forall\bar{a},\bar{b}.D \Rightarrow t \arrow T~\bar{a} \sgap
          \bar{b} \cap \bar{a} = \emptyset \sgap        
          p : [\bar{t}/\bar{a}] t \turns \forall \bar{b'}.(D' \gd \tenv_p)}
         {K~p : T~\bar{t} \turns \forall \bar{b'},\bar{b}.(D'\wedge [\bar{t}/\bar{a}] D \gd \tenv_p)}
\eda
\vspace{-0.4cm}
}

In this section, we define the set of well-typed expressions.
\bda{llcl}
 \mbox{Expressions} & e & ::= & K \mid x \mid \lambda x.e \mid e~e \mid (e::C\Rightarrow t)
                \mid \REC f\, \IN\ e \mid
                  \CASE\ e~ \OF\ [p_i \arrow e_i]_{i\in I}
\\ \mbox{Patterns} & p & ::= & x \mid (p,p) \mid K~p  ~~~~~~  \mbox{Types} ~~~~~~~~~~~~~ t ~~ ::= ~~ a \mid t \arrow t \mid T~\bar{t}
\\ \mbox{Constraints} & C & ::= & t=t \mid C \wedge C ~~~~~~~~  \mbox{Type Schemes} ~~ \sigma ~~ ::= ~~ t \mid \forall \alphavec.C \Rightarrow t
\eda
For simplicity, we omit let-definitions but may make use of them in examples.
We consider pattern matching syntax as syntactic sugar for case expressions.
GRDT definitions have been preprocessed and are recorded in some initial type environment.
E.g., we find that $IsZero:\forall a.a=Int \Rightarrow Exp~a \in \tenv_{init}$ etc.~for 
the GRDT from Example~\ref{ex:eval}.

In Figure~\ref{f:type-system} we define the set of well-typed GRDT programs
in terms of typing judgments $C, \tenv \turns e : t$. 
Rules \tlabel{Abs}, \tlabel{App} and \tlabel{Rec} are standard.
In rule \tlabel{Eq} the side condition $C \implies t_1=t_2$ holds iff
(1) $C$ does not have a unifier, or (2) for any unifier $\phi$ of $C$ we have
that $\phi(t_1)=\phi(t_2)$ holds. Hence, we can change the type of an expression
given some appropriate type assumptions.
In rule \tlabel{Var-x} we build a type instance of a type scheme.
Rule \tlabel{Case} is standard again.
Rule \tlabel{Annot} deals with type annotation. Note that we only allow
for {\em closed} type annotations, i.e.~the set of variables appearing in the type and constraint
component is assumed to be universally bound. 
We consider this is a non-essential restriction and leave the
extension to ``open'' annotations for future work.
W.l.o.g., we assume that there are no name clashes
with other variables in the typing judgment.
Rule \tlabel{Pat} is interesting. We type the body of a pattern clause under
the additional constraints arising out of the pattern.
Note that we make use of an 
auxiliary judgment $p:t \turns \forall \bar{b}.(D \gd
\tenv_p)$ which establishes a relation among pattern $p$ of type
$t$ and the binding $\tenv_p$ of variables in $p$. 
Variables
$\bar{b}$ refer to all ``existential'' variables. 
Logically, these variables must be considered as universally quantified.
Hence, we write $\forall \bar{b}$.
The side condition $\bar{b}\cap fv(C,\tenv,t_2)=\emptyset$ prevents
existential variables from escaping.
In rule
\tlabel{Pat-Pair}, we assume that there are no name clashes
between variables $\bar{b_1}$ and $\bar{b_2}$.
Constraint $D$ arises from constructor occurrences in $p$.

In contrast to standard Hindley/Milner, the GRDT system as presented does not enjoy principal types.

\begin{example} \label{ex:no-pt}
We assume a primitive operation {\tt (+)::Int->Int->Int}.
\vspace{-1mm}
\begin{code}
data Erk a = (a=Int) => I a | (a=Bool) => B a 
f x y = case x of I z -> y+z
\end{code}
\vspace{-1mm}
%
%
We find that $\forall a. Erk~a \arrow Int \arrow Int$, $\forall a.Erk~a \arrow a \arrow Int$,
$\forall a. Erk~a \arrow a \arrow a$, $\forall b,c.Erk~(Int\arrow Int) \arrow b \arrow c$,  
$\forall b,c.Erk~(Bool \arrow Bool) \arrow b \arrow c$, ...,
are all incomparable types but there does not seem to be a most general type. 
Note that the last set of types is correct.  We temporarily make use of
$False$ (which is equivalent to e.g.~$Int=Int \arrow Int$) under which we
can give any type to the body of the case expression.  As pointed out
in~\cite{first-class-phantom} such ``meaningless'' types can safely be
omitted. In essence, the program text belonging to a meaningless type
represents ``dead code'' and can always be replaced by $\bot:\forall a.a$.
Note that meaningless types will always destroy the principal types
property.  Hence, we will rule out such types by strengthening rule
\tlabel{Eq}. We drop the first condition and only impose the second
condition that for any unifier $\phi$ of $C$ we have that
$\phi(t_1)=\phi(t_2)$ holds.  Note that the first three types are
``meaningful'' but there is still no most general type.
\end{example}

\vspace{-1mm}
The potential loss of principal types for GRDTs has already been observed by
Cheney and Hinze~\cite{first-class-phantom}.
As a solution they suggest explicitly
providing  result-type annotations for case expressions. However, the above example shows 
that this is not sufficient to retain principal types.
As shown by Simonet and Pottier~\cite{simonet-pottier-hmg}, we can trivially achieve principal types by
enriching the set of constraints allowed to appear in typing judgments.
E.g., we can give {\tt f} the non-expressible ``principal type''
$\forall a, t_y, t. (a=Int \implies (t_y=Int\wedge t=Int)) 
   \Rightarrow Erk~a \arrow t_y \arrow t$. Notice the use of 
 Boolean implication ($\implies$) 
to describe the set of types which can be given to {\tt f}.
There are several good reasons 
why we do not want to admit such expressive types.
For example, type inference becomes more complex, 
and types become less readable.


\mysection{Efficient Type Inference} \label{sec:inference}

\fig{f:type-inference}{Generating Constraints}{

\bda{cc} 

\ba{c}
\ba{cc} \tlabel{Var-x} & \myirule{(x:\forall\bar{a}.C\Rightarrow t)\in\tenv}{\tenv,x \trl (C \gd t)} \ea 
\ea
 &
\ba{cc}
  \tlabel{Rec} &
  \myirule{\tenv.f:t_1, e \trl (F \gd t_2) \sgap \mbox{$t_1$ fresh}}
          {\tenv, \REC f\, \IN\ e \trl (F, t_1=t_2 \gd t_1)}
\ea
\eda
\bda{cc}
\ba{cc}
 \tlabel{App} &
  \myirule{\tenv, e_1 \trl (F_1 \gd t_1) \\
           \tenv, e_2 \trl (F_2 \gd t_2) \\
           \mbox{$t$ fresh} \sgap
           F \equiv F_1 \wedge F_2 \wedge t_1=t_2\arrow t }
        {\tenv, e_1~e_2 \trl (F \gd t)}
\ea 
 &
\ba{cc} \tlabel{Abs} &
   \myirule{\mbox{$a$ fresh} \\
            \tenv.x:a, e \trl (F \gd t)}
           {\tenv, \lambda x.e \trl (F \gd a\arrow t)}
\ea
 \eda

\bda{cc}
 \tlabel{Case} &
 \myirule{ \tenv, p_i \arrow e_i \trl (F_i \gd t'_i) \sgap \mbox{for $i\in I$}  \sgap
         \tenv, e \trl (F_e \gd t_e) \sgap \mbox{$t_1, t_2$ fresh} \\ 
         F \equiv F_e \wedge t_1=t_e \arrow t_2 \wedge 
           \bigwedge_{i\in I} (F_i \wedge t_1=t'_i)}
         {\tenv, \CASE\ e~ \OF\ [p_i \arrow  e_i]_{i\in I} \trl (F \gd t_2)}
\\ \\
 \tlabel{Pat} &
 \myirule{p \turns \forall \bar{b}.(D \gd \tenv_p \gd t_1) \sgap
          \tenv \cup \tenv_p, e \trl (F_e \gd t_e)  \sgap \mbox{$t$ fresh} \\
           F \equiv \forall \bar{b}. (D \implies \bar{\exists}_{\tv(\tenv,\bar{b},t_e)}.F_e) \wedge 
                t=t_1\arrow t_e }
         {\tenv, p \arrow e \trl (F \gd t)}
\eda
\bda{cc}
 \ba{cc}
   \tlabel{Annot} &
   \myirule{\tenv, e \turns (F' \gd t') \sgap \bar{a}=\tv(C,t) \\
            F \equiv \forall \bar{a}. ((C \wedge t=t) \implies \bar{\exists}_{\tv(\tenv,t')}.F')}
          { \tenv, (e::(C\Rightarrow t)) \trl (F \gd t')}
 \ea
 \ba{cc}
   \tlabel{Pat-Var} &
   \myirule{\mbox{$t$ fresh}}
         {x \turns (True \gd \{x:t\} \gd t)}
\ea
\eda
\bda{cc}
 \tlabel{Pat-Pair} &
  \myirule{p_1 \turns \forall \overline{b_1}.(D_1 \gd \tenv_{p_1} \gd t_1) \sgap
           p_2  \turns \forall \overline{b_2}.(D_2 \gd \tenv_{p_2} \gd t_2)}
          {(p_1,p_2)  \turns  
      \forall \overline{b_1,b_2}. (D_1 \wedge D_2 \gd \tenv_{p_1}\cup \tenv_{p_1} \gd (t_1,t_2))}
\\ \\
 \tlabel{Pat-K} &
 \myirule{K:\forall\bar{a},\bar{b}.D \Rightarrow t \arrow T~\bar{a} \sgap
          p \turns \forall \bar{b'}.(D' \gd \tenv_p \gd t_p) \sgap
          \mbox{$\phi$ m.g.u.~of $t_p=t$}}
         {K~p \turns \forall \bar{b'},\bar{b}.(\phi(D')\wedge D \gd \phi(\tenv_p) \gd T~\phi(\bar{a}))}
\eda
}

We introduce an efficient inference method for GRDTs which is divided into two steps.
In a first step, we take the standard route and generate an appropriate set of constraints out of the program text.
For this purpose, we assume an enriched constraint language consisting of Boolean connectives such as $\implies$ (implication)
and quantifiers $\forall$ and $\exists$.
If necessary we refer to ``simple'' constraints as the set of constraints admitted in the type system
described in the previous section.
In the second step, we perform some equivalence transformations on constraints such that resulting constraints
can be solved efficiently by unification under a mixed prefix~\cite{147067}.

In Figure~\ref{f:type-inference}, we describe the constraint generation rules
in terms of judgments $\tenv, e \trl (F \gd t)$. We commonly refer to $F$ as the {\em inferred} constraint.
Notice the use of Boolean implication ($\implies$) and universal quantification ($\forall$).
In rule \tlabel{Pat}, we use $\bar{\exists}_V.F$ as a short-hand for $\exists\tv(F)-V.F$.

\vspace{-1mm}
\begin{example} \label{ex:no-pt2}
Consider constraint generation for Example~\ref{ex:no-pt} where 
$\tenv_{init} = \{ I: \forall a. a=Int \Rightarrow a \arrow Erk~a, B: \forall a. a=Bool \Rightarrow a \arrow Erk~a \}$.
Let $e\equiv\lambda.x. \lambda y.\CASE\ x~ \OF\ I~z \arrow (y+z)$ (desugared version of {\tt f}'s program text).
Then $\emptyset, e \trl ( t_x=Erk~a \wedge (a=Int \implies (t_y=Int\wedge t_1=Int)) 
\gd t_x \arrow t_y \arrow t_1)$.
Note that we have slightly simplified the constraint and type.
Often, we ``normalize'' the resulting type and constraint and write
\bda{l}
 t = t_x \arrow t_y \arrow t_1, t_x=Erk~a,
(a=Int \implies (t_y=Int, t_1=Int))
\eda
where $t$ refers to the type of expression $e$.
\end{example}
\vspace{-1mm}

The important observation is that 
based on the following first-order equivalences we can normalize constraints:
(1) $(F_1 \implies Q a.F_2) \leftrightarrow Q a.(F_1 \implies F_2)$
where $a\not\in\tv(F_1)$ and (2) $(Q a.F_1) \wedge 
(Q b.F_2) \leftrightarrow Q a,b.(F_1 \wedge F_2)$
where $a \not\in\tv(F_2)$, $b\not\in\tv(F_1)$ and $Q \in \{ \exists, \forall \}$ and (3) $C_1 \implies (C_2 \implies C_3) 
\leftrightarrow (C_1 \wedge C_2)\implies C_3$.
We exhaustively apply the above identities from left to right.
W.l.o.g., we assume that bound variables have been renamed.
We can conclude that each
inferred constraint $F$ can be equivalently represented
as ${\cal Q}.
C_0 \wedge (D_1 \implies C_1) \wedge ... \wedge (D_n \implies C_n)$
where $C_0$, $D_1$, $C_1$,...,$D_n$,$C_n$ are constraints
and ${\cal Q}$ is a mixed-prefix of quantifiers of the form
$\forall\overline{a_1}.\exists\overline{b_1}...\forall\overline{a_n}.\exists\overline{b_n}$.
Commonly, we refer to the last constraint as the {\em normalization} of $F$.
Normalized constraints can be efficiently solved by unification under a
mixed prefix  as follow:
(1) Build an m.g.u.~$\phi$ of $C_0$ under prefix {\cal Q} (see~\cite{147067}
for details on unification under a mixed prefix), and
(2) set $E_i = \phi_i\circ \phi(C_i)$ if m.g.u.~$\phi_i$ 
of $\phi(D_i)$ under prefix {\cal Q} exists,
or $E_i = True$ otherwise for $i=1,...,n$.
(3) Build the m.g.u.~$\psi$ of $C_0 \wedge E_1 \wedge ... \wedge E_n$ under
prefix {\cal Q}.
In case all three steps were successful, we write $\psi={\it solve}(F)$.
Note that in such a situation, we find that 
$\psi$ is a {\em solution} of $F$, i.e. $\models \psi(F)$ holds.
We write $F_1 \models F_2$ to denote that
any model of $F_1$ is a model of $F_2$. $F_1$ is commonly omitted if $True$. 

\vspace{-1mm}
\begin{example}
Consider the constraint generated in Example~\ref{ex:no-pt2}.
In solving step (2), we generate $t = t_x \arrow t_y \arrow t_1, t_x=Erk~a, (t_y=Int, t_1=Int)$.
Hence, we find the solution $\psi=[Int/t_y, Int/t_1, Erk~a \arrow Int \arrow Int /t]$.
Hence, expression $\lambda.x. \lambda y.\CASE\ x~ \OF\ I~z \arrow (y+z)$
can be given type $\forall a. Erk~a \arrow Int \arrow Int$.
\end{example}
\vspace{-1mm}

Note that the inferred type is not principal. See the discussion in the previous section.
Hence, the question is whether this type is acceptable.
We will address such issues and how to check for principality in Section~\ref{sec:principal}.
The least we can state at this stage is that our inference method is sound.

\vspace{-1mm}
\begin{theorem}[Soundness of Inference] \label{th:trl-sound}
Let $\tenv$ an environment, $e$ an expression,
$F$ a constraint and $t$ a type such that $\tenv, e \trl (F \gd t)$.
Let $\psi={\it solve}(F)$.
Then $True, \psi(\tenv) \turns e : \psi(t)$.
\end{theorem}
\vspace{-1mm}

There are cases where our method fails, although the program is well-typed.

\vspace{-1mm}
\begin{example} \label{ex:size}
Here is an example taken from~\cite{first-class-phantom}.
\vspace{-1mm}
\begin{code}
data R a = (a=Int) => RInt | forall b c.(a=(b,c)) => RProd (R a) (R b)
size RInt = 1
size (RProd a b) = (size a) + (size b)
\end{code}
\vspace{-1mm}
We generate the (simplified) constraint
$t=R~a \arrow t_1, (a=Int \implies t_1=Int),
\forall b, c.(a=(b,c) \implies t=R~b \arrow t_2, t=R~c \arrow t_3, t_1=t_2,t_1=t_3,t_1=Int)$.
Normalization yields
$(\forall b,c.(t=R~a \arrow t_1, (a=Int \implies t_1=Int),
 (a=(b,c) \implies t=R~b \arrow t_2, t=R~c \arrow t_3, t_1=t_2,t_1=t_3,t_1=Int))$.\footnote{We silently
drop the outermost ``empty'' forall quantifier and the existential quantifier over $t$.}
In the solving step (2), we generate
$t=R~a \arrow t_1,t_1=Int, R~(b,c)=R~b, R~(b,c)=R~c, t_1=t_2,t_1=t_3,t_1=Int$
which cannot be solved by unification under the prefix $\forall b,c$.
However, {\tt size} is well-typed under type $\forall a.R~a \arrow Int$.
\end{example}
\vspace{-1mm}

\vspace{-1mm}
\begin{example} \label{ex:erk}
We consider a variation of Example~\ref{ex:no-pt}. Additionally, we make use of
a primitive operation {\tt (\&\&)::Bool->Bool->Bool}.
\vspace{-1mm}
\begin{code}
f (I x) = x + 1
f (B x) = x && True
\end{code}
\vspace{-1mm}
We generate $t=Erk~a \arrow t_1, (a=Int \implies t_1=Int), (a=Bool \implies t_1=Bool)$.
In solving step (2), we generate
$t=Erk~a \arrow t_1, t_1=Int, t_1=Bool$ which is not solvable. 
Hence, our inference method fails. On the other hand, {\tt f} can be given type $\forall a.Erk~a \arrow a$.
\end{example}
\vspace{-1mm}

We draw the following conclusions.
Our inference method may fail because GRDT programs often make 
use of polymorphic recursion (see Example~\ref{ex:size}).
Another reason for failure is that we naively combine the inference results from different branches
(see Example~\ref{ex:erk}).
Indeed, other inference approaches~\cite{simonet-pottier-hmg,gad} face the same problem.
Hence, we will need to sufficiently annotate the program such that inference succeeds.
It should be clear that we must provide types for polymorphic recursive
functions. The problem is shown to be undecidable for Hindley/Milner~\cite{henglein}.
However, instead of providing full annotation we would like
to provide only a minimal amount of information.
E.g., in case of the {\tt size} function it is sufficient to provide only information
about the input type.

\vspace{-1mm}
\begin{example} \label{ex:size-refined}
Recall Example~\ref{ex:size}.
We guess that {\tt size} must take in values of type $R~a \arrow b$ for any $a$ and for some $b$.
The programmer could indicate this information via ``partial'' annotations of the
form {\tt size::R a->\_}. For type inference purposes, we simply assume that
{\tt size} has type $\forall a, b. R~a \arrow b$.
Under this assumption, we generate the (simplified) constraint
$t=R~a \arrow t_1, (a=Int \implies t_1=Int),
\forall b, c.(a=(b,c) \implies t_1=Int)$.
Our solving method succeeds here and yields that {\tt size} has type $\forall a. R~a \arrow Int$.
\end{example}
\vspace{-1mm}

In general, we propose the following refinement of rule \tlabel{Rec}.

\vspace{-4mm}
\bda{cc}
  \tlabel{Rec-Guess} &
  \myirule{\tenv.f:\sigma, e \trl (F \gd t_2) \sgap \mbox{guess a type $\sigma$} \sgap
           \psi={\it solve}(F) \\ \bar{a}=\tv(\psi(t_2))-\tv(\psi(\tenv))
           \sgap \psi(\tenv).f:\forall\bar{a}.\psi(t_2), e \trl (F' \gd t_2')}
          {\tenv, \REC f\, \IN\ e \trl (F, \forall\bar{a}.(\psi(t_2)=t_2' \implies F') \gd t_2)}
\eda
Note that we also have to check that the result we obtain from guessing is indeed a valid type.
That is, we first build the type $\sigma'=\forall\bar{a}.\psi(t_2)$ and perform
inference again. Then, we verify that the type inferred, represented by $(F' \gd t_2')$,
under assumption $f:\sigma'$ subsumes $\sigma'$.
The constraint $\forall\bar{a}.(\psi(t_2)=t_2' \implies F')$ guarantees that this condition holds.
Note that a similar idea has been mentioned in~\cite{696195}.

Obviously, this refined method requires that we have a good heuristic for guessing types.
We argue that in many cases we can guess from the program text alone
which
``input'', i..e.~lambda-bound, variables are connected to GRDTs.
See Examples~\ref{ex:eval} and \ref{ex:size}. 
However, the upcoming Example~\ref{ex:grdt-ex2} shows that this is not necessarily the case.
Lambda-bound variables may be connected via type constraints to GRDTs. 
For such cases, we simply introduce a fresh universal variable.  Note that
we can further refine our method by performing a couple of iterations.  In
particular, this helps if $\forall a.a$ is our initial guess.  

\comment{
\ms{if this
condition is still not satisfied we could iterate again and
  again ...}\pjs{True but when do you give up?}
\ms{well, it's an undecidable problem in general. it should be fine just to say that there's clearly a limit
of what can be inferred and what must be annotated.}

\ms{we should argue that the program text indicates if a function is recursive and at which position
we expect a GRDT. Otherwise, somebody might argue just simply guess random types and perform inference.
Guessing that the type must be a GRDT is easy.
Check that our heuristic in combination with further coming refinements
is successful for most GRDT examples.}\pjs{I agree at the moment the whole
inference process is too up in the air, we only want to look for GRDT
solutions where they can possible be!}
}

Note that the refined method will not succeed in case of Example~\ref{ex:eval}
(if we guess that {\tt eval} has type $\forall a,b. Exp~a \arrow b$).
The problem here is that the type changes for each branch (the same happens in Example~\ref{ex:erk}).
Our inference method still naively combines the results from different branches.
Hence, we fail.
Further refinements of our inference scheme are necessary.
In Section~\ref{sec:enumerate}, we show how to build solutions {\em automatically}
by inspecting sub-results of inference.
In the next section, we first
establish a criteria under which constraint solving always succeed.
Failure of the criteria may prove helpful to assist {\em user-guided} 
input in terms of type annotations such that inference succeeds eventually.


\mysection{Constraint Solving Criteria} \label{sec:inf-criteria}

The observations in the previous section let us conclude that 
inference may fail because types change in different branches
(assuming that we exclude the event of a type error).
We are looking for a sufficient criteria under which we can guarantee that inference will succeed.
In case the criteria cannot be satisfied, the hope is that we obtain some crucial information
to identify which type information is missing such that inference might succeed.
Our task is to identify all type equations arising out of different branches which may lead
to some inconsistencies. Looking at this question from a different angle, we need to
identify which types must be known such that no inconsistency will arise.
For this purpose, we keep track of types which are ``known''.
We introduce a predicate $known(t)$ which states that type $t$ is known.
E.g., type $t$ is given through an annotation. However, we may also implicitly propagate
known types. E.g., assume inference generates the constraint $t=(t_1,t_2)$
then we conclude that also $t_1$ and $t_2$ are known. We can capture this via the following relations.
\begin{equation} \label{eq:known-relations}
\ba{l}
\forall t_1,t_2. (known(t_1 \arrow t_2) \leftrightarrow known(t_1) \wedge known(t_2)) \\
 \forall t_1,t_2. (known((t_1,t_2)) \leftrightarrow known(t_1) \wedge known(t_2)) 
\ea
\end{equation}
Note that a type must be known if the different branches disagree.
Assume $E_t$ denotes the equations constraining $t$ from a particular branch.
Then, the constraint $known(t) \vee E_t$ expresses the fact that $t$ is known or
the constraints in $E_t$ will become effective.
Let's focus on two branches and observe the effect on $t$. We find the constraint
$(known(t) \vee E_t) \wedge (known(t) \vee E'_t)$
which is equivalent to $(known(t) \vee (E_t \wedge E'_t))$ (2).
Assume the two branches have the same effect on $t$. E.g., this is the case
for $t_1$ in case of Example~\ref{ex:size-refined}.
Then, (2) is equivalent to $E_t,E'_t$ indicating that $t_1$ must not be known necessarily.
On the other hand, in case of Example~\ref{ex:erk} the branches disagree.
Hence, (2) is equivalent to $known(t_1)$ indicating that $t_1$ must be known.

We incorporate this idea of identifying which constraints must be known
into our constraint generation rules.
We adapt rule \tlabel{Pat} from Figure~\ref{f:type-inference} as follow:
\vspace{-1.5mm}
\bda{cc}
 \tlabel{Pat} &
 \myirule{p \turns \forall \bar{b}.(D \gd \tenv_p \gd t_1) \sgap
          \tenv \cup \tenv_p, e \trl (F_e \gd t_e)  \sgap \mbox{$t$ fresh} \sgap \bar{a}=\tv(F_e) \\
         \ba{lcl}
           F & \equiv & \forall \bar{b}. ((D \implies \bar{\exists}_{\tv(\tenv,\bar{b},t_e)}.F_e) \wedge t=t_1\arrow t_e) \wedge \\
           &&      \bigwedge_{a \in \bar{a}} (known(a) \vee 
                    (\exists\tv(D,F_e)-\tv(\tenv,a,\bar{b}).(D \wedge F_e)))
         \ea }
         {\tenv, p \arrow e \trl (F \gd t)}

\vspace{-2mm}
\eda
For simplicity, we only consider expressions $e$ which do not contain nested case expressions, hence, $F_e$ is a simple constraint.
Otherwise, we will need to manipulate the program text by introducing auxiliary (local) function definitions
and ``flattening'' the program by performing lambda-lifting.

In addition to the existing normalization steps we make use of the following identities:
$(F_1 \vee F_2) \wedge (F_1 \vee F_3) \leftrightarrow (F_1 \vee (F_2 \wedge F_3))$
and $(F_1 \vee \exists a.F_2) \leftrightarrow \exists a.(F_1 \vee F_2)$ where $a\not\in \tv(F_1)$.
Hence, the constraint resulting out of $e$ is now of the form
${\cal Q}.C_0 \wedge (D_1 \implies C_1) \wedge ... \wedge (D_n \implies C_n) \wedge K$
where $C_0$, $D_i$ and $C_i$ consist of conjunction of equations
and $K$ is equivalent to $\bigwedge_{\bar{a}} (known(a) \vee E_a)$ where $E_a$ is a conjunction of equations
constraining $a$.

The main point of our formulation is that 
we now can query the normalized constraint to identify which types must be known.

\vspace{-1mm}
\begin{example}
Here is a variation of Example~\ref{ex:erk} where we make use of {\tt (>)::Int->Int->Bool}.
\vspace{-3mm}
\begin{code}
h (I x) = x > 1
h (B x) = x && True
\end{code}
\vspace{-1mm}
The normalized constraint generated out of the program text is as follow.
We denote this constraint by $F$.
\bda{ll}
 t=Erk~a \arrow t_1 \wedge (known(t_1) \vee (t_1=Bool \wedge t_1=Bool) & (C_0 \wedge K) \\
 (a=Int \implies t_1=Bool) \wedge (a=Bool \implies t_1=Bool)) & ((D_1 \implies C_1) \wedge(D_2 \implies C_2)) \\
\eda
We find that $F \not \models known(t_1)$. That is, $t_1$ need 
not be known.
Hence, we can safely combine the results from different branches.
Hence, inference succeeds. 
Indeed, we infer that {\tt h} has type $\forall a.Erk~a \arrow Bool$.
Note that a similar reasoning applies to Example~\ref{ex:size-refined}.
\end{example}
\vspace{-1mm}

More formally, we define 
${\cal K}(F) = \{ known(a) \mid F \models known(a) \}$.
Silently, we assume that the known relations described in (\ref{eq:known-relations}) are always included.
Note that for a given $a$ we can decide $F \models known(a)$ by 
putting $F, \neg known(t)$ 
into clause form and test for a contradiction by applying resolution. 
Note that resolution is complete for refutation (see e.g.~\cite{Sho67}). 
Hence, we have a decidable check to verify if inference is successful.

\vspace{-1.5mm}
\begin{lemma} [Constraint Solving Criteria] \label{le:suff-known-lemma}
Let $e$ be an expression containing no annotations and no nested case expressions.
Let ${\cal Q}.C_0 \wedge (D_1 \implies C_1) \wedge ... \wedge (D_n \implies C_n) \wedge K$
be the (normalized) constraint generated and $t$ the type of $e$.
Let $U$ be a (simple) user-provided constraint where $K_U = \wedge_{a\in\tv(U)} known(a)$.
If ${\cal K}(U \wedge K_U \wedge {\cal Q}.C_0 \wedge (D_1 \implies C_1) \wedge ... \wedge (D_n \implies C_n) \wedge K)=
      {\cal K}(U \wedge K_U \wedge {\cal Q}.C_0 \wedge (D_1 \implies C_1) \wedge ... \wedge (D_n \implies C_n))$
and $U\wedge {\cal Q}.C_0 \wedge (D_1 \implies C_1) \wedge ... \wedge (D_n \implies C_n)$ is satisfiable,
then $U\wedge {\cal Q}.C_0 \wedge (D_1 \implies C_1) \wedge ... \wedge (D_n \implies C_n)$
has a solution.
\end{lemma}
\vspace{-1.5mm}

The above lemma suggests the following strategy.
By default always perform efficient solved form inference.
In case we fail, pick a $t$ and check whether $t$ must be known.
The question of guessing an appropriate $t$ is non-trivial.
The ``shape'' of $t$ is constrained by the variables and equations generated.
Hence, there is only a finite number of non-trivial
$known(t)$. 
However, enumerating all possibilities might be infeasible in practice.
A good guess might be to consider all variables involved in a minimal unsatisfiable subset (e.g.~\cite{interactive}) of constraints in
$C_0 \wedge E_1 \wedge ... \wedge E_n$.

\vspace{-1mm}
\begin{example}
Recall Example~\ref{ex:erk}.
In an intermediate step, we attempt to solve
$t=Erk~a \arrow t_1, t_1=Int, t_1=Bool$ which fails.
We find that $t_1=Int, t_1=Bool$ form a minimal unsatisfiable subset. We pick a variable from this set (there's
only one here).
The (normalized) constraint generated via the ``known'' inference approach is as follow.
$t=Erk~a \arrow t_1, (a=Int \implies t_1=Int), (a=Bool \implies t_1=Bool), (known(t_1) \vee (t_1=Int,t_1=Bool)$.
Immediately, we find that $known(t_1)$ is a logical consequence.
The user did not provide any information about $t_1$, hence, we conclude that $t_1$ must be provided
such that inference succeeds.
E.g., we find that if the user provides $t_1=a$ the conditions of the above lemma are fulfilled.
Indeed, efficient inference succeeds now.
\end{example}
\vspace{-1mm}

A similar reasoning applies to Example~\ref{ex:eval}.
Here is another interesting example.

\vspace{-1mm}
\begin{example} \label{ex:grdt-ex2}
Consider the following program where we make use primitive functions
{\tt h1::Erk Int->Int->Int},
{\tt h2::Erk Bool->Bool->Int} and
{\tt h3::a->a->Bool}.
\vspace{-1mm}
\begin{verbatim}
f = \y -> \x -> (h3 y x, -- (1)
                 case y of  I z -> h1 x z    -- (2)
                            B z -> h2 x z)   -- (3)
\end{verbatim}
\vspace{-1mm}
We generate the following constraint 
\bda{ll}
 t=t_y \arrow t_x \arrow t_1, t_1=(t_2,t_3,t_4), & (0)
\\ t_y=t_x, t_2=Bool,  & (1)
\\ (t_z=Int \implies t_z=Int,t_x=Erk~Int,t_3=Int), & (2)
\\ (t_z=Bool \implies t_z=Bool, t_x=Erk~Bool, t_4=Bool), & (3)
\\ (known(t_x) \vee (t_x =Erk~Int, t_x=Erk~Bool)
\eda
Note that no user annotations are provided
and the type of {\tt x} changes. We have seen previously that this is may make
our inference method fail (see Examples~\ref{ex:eval} and~\ref{ex:erk}).
However, efficient inference succeeds, i.e.~solving of constraints (0-3) yields a solution, and we can formally show why.
We can argue that the type of {\tt y} is known because the case expression forces {\tt y} to be
a GRDT~$Erk~a$. Hence, we add the fact that $known(t_y)$. In combination with constraint (1)
we can establish that the assumptions of Lemma~\ref{le:suff-known-lemma} are satisfied.
\end{example}
\vspace{-1mm}



\mysection{Incremental Building of Solutions} \label{sec:enumerate}

Instead of immediately solving constraints generated by 
Figure~\ref{f:type-inference}, or
in case of failure trying to find which types must be known as suggested in Section~\ref{sec:inf-criteria},
we show how to build solutions incrementally.
We illustrate our approach by example first.

\vspace{-1mm}
\begin{example} \label{ex:enumerate}
Consider a variation of Example~\ref{ex:erk}.
\bp
data Erk a = (a=Int) => I a | (a=Bool) => B a \\
h = $\lambda$x.$\lambda$y. \= case x of \= I z -> z + y 
    \\ \> \>  B z -> z \&\& y
\ep
We generate the following constraint.
\bda{ll}
 t=Erk~a \arrow t_y \arrow t_r \wedge~ & (C_0) \\
 (a=Int \implies (t_y=Int \wedge t_r=Int)) & (D_1 \implies C_1) \\
 (a=Bool \implies (t_y =Bool \wedge t_r=Bool)) & (D_2 \implies C_2)
\eda
%
Note that inference fails here.
Instead, for each $C_0 \wedge D\implies C$ we calculate ${\cal S}=\{ E \mid
C_0 \wedge D \wedge C \implies E \}$ where $E$ is a conjunction of equations,
i.e.~the set of all implied equations which 
potentially take part in a solution.
We find that
\bda{lcl}
 {\cal S}_1 & = & \{  \fset{t_y=Int}, \fset{t_y=a},   
                 \fset{t_r=Int}, \fset{t_r=a}, \\
            && \fset{t_y=Int, t_r=Int}, \fset{t_y=Int, t_r=a}, \\ 
            && \fset{t_y=a, t_r=Int}, \fset{t_y=a, t_r=a} \} \\
 {\cal S}_2 & = & \{  \fset{t_y=Bool}, \fset{t_y=a},   
                \fset{t_r=Bool}, \fset{t_r=a}, \\
            && \fset{t_y=Bool, t_r=Bool}, \fset{t_y=Bool, t_r=a}, \\ 
            && \fset{t_y=a, t_r=Bool}, \fset{t_y=a, t_r=a} \}
\eda
Then, we go through all combinations
$S_1 \in {\cal S}_1$ and $S_2 \in {\cal S}_2$ 
to find a solution. Note that there can only be a finite number of combinations.
E.g., $S= \{ t_y=a, t_r=a \}$ is such a solution.
As we will see later, this solution is even principal.
Hence, {\tt h} has the
principal type $\forall a. Erk~a \arrow a \arrow a$.
\end{example}
\vspace{-1mm}

Note that the above method applied to Example~\ref{ex:eval} would e.g.~infer
the type $\forall a.Erk~a \arrow a$.
The important observation is that for
satisfiable equations under a mixed prefix we can enumerate all implied equations.
We define 
${\cal S}(F)= \{ E \mid F \implies E, \mbox{$E$ consists of equations only} \}$.

\vspace{-1mm}
\begin{lemma}[Finite Solutions] \label{le:finite-solutions}
Let ${\cal Q}.C$ be a satisfiable set of equations under a mixed prefix {\cal Q}. Then,
${\cal S}({\cal Q}.C)$ 
is finite (assuming a canonical form of equations) 
and each element consists 
of only a finite number of equations.
\end{lemma}
\vspace{-1mm}

The following lemma shows that we can construct a solution out of the implied constraints resulting
from the different branches if the solution space is non-trivial (i.e.~does not only contain $False$).
For convenience, we define
$E_{\psi}= \{ a=\psi(a) \mid a\in {\it domain}(\psi) \}$
to be the constraint representation of a substitution $\psi$.

\vspace{-1mm}
\begin{lemma} [Building Solutions] \label{le:enum-solutions}
Let ${\cal Q}.C_0 \wedge (D_1 \implies C_1) \wedge ... \wedge (D_n \implies C_n)$ be
such that $\psi$ is a solution and
${\cal Q}.C_0 \wedge D_i \wedge C_i$ is satisfiable for $i=1,..,n$.
Then, there exist $S_i \in {\cal S}({\cal Q}.C_0 \wedge D_i \wedge C_i)$ for $i=1,..,n$
such that $E_{\psi}$ and $\bigwedge_{i=1,...,n} S_i$ are equivalent w.r.t. $\tv({\cal Q}.F)$.
\end{lemma}
\vspace{-1mm}


\mysection{Principal Types} \label{sec:principal}

In Example~\ref{ex:no-pt} we have observed that the GRDT system does not enjoy principal types
in general. Given the complexity of type inference for GRDTs we are quite content to infer
{\em a} type. However, if possible we would like to report to the user if a type is {\em not} principal.
In this section, we identify a
a necessary criteria for a type to be principal.
Hence, we obtain an efficient but incomplete procedure for testing if
a type is not principal.
Based on the enumeration technique given in the previous
section we can even find a principal type.
We simply consider all combinations of possible solutions and check if there is a principal solution.

First, we define principal solutions.
We say $\psi$ is a {\em principal solution} of $F$ iff $\models \psi(F)$ and given
another solution $\phi$ of $F$ we have that $\exists \theta. \phi = \theta \circ \psi$.
That is the substitution $\psi$ is more general than any other solution.
It is easy to show that every principal solution yields a principal type.
Every principal solution must satisfy the following criteria.

\vspace{-1mm}
\begin{lemma} [Necessary Principal Solution Criteria] \label{le:necessasry-criteria}
Let $\psi$ be a principal solution of $F$.
Let $F'$ be the skolemized version of $F$ of the form $C_0,(D_1 \implies C_1),...,(D_n \implies C_n)$.
Then, 
$(E_{\psi},C_0,\bigwedge_{i=1,...,n} D_i) \leftrightarrow 
      (C_0, \bigwedge_{i=1,...,n} (C_i, D_i))$ where 
$E_{\phi}= \{ a=\phi(a) \mid a\in {\it domain}(\psi) \}$.
\end{lemma}
\vspace{-1mm}

An interesting observation is that ``meaningless'' types are never principal.
\vspace{-1mm}
\begin{example}
Recall  the constraint generated out of {\tt f}'s program text (see Example~\ref{ex:no-pt2})
\bda{l}
 t = t_x \arrow t_y \arrow t_1, t_x=Erk~a,
(a=Int \implies (t_y=Int, t_1=Int))
\eda
The meaningless  type $\forall b,c.Erk~(Int\arrow Int) \arrow b \arrow c$ from Example~\ref{ex:no-pt}
corresponds to the solution $\psi=[Erk~(Int\arrow Int) \arrow b \arrow c/t]$.
We omit skolemization which is unnecessary here.
We find that the lhs of the logical condition 
is unsatisfiable (since $\phi(a) = Int \arrow Int$) 
whereas the rhs is. 
Hence, $\psi$ is not principal.
\end{example}
\vspace{-1mm}

The above applies to all meaningless types.
A silent assumption is that 
constraints appearing
in the types of GRDT constructors $K$ are always satisfiable.
Furthermore, we need to rule out the case that the program is free of type errors.\footnote{Remember
that with a meaningless type we can even type ill-typed programs because under the $False$ assumption
we can give any type to an expression.}

Unfortunately, our (necessary) principal types condition seems to weak in practice 
to identify non-principal types.
In Example~\ref{ex:no-pt} we argued that {\tt f} has no principal type.
However, we find that types  
$\forall a. Erk~a \arrow Int \arrow Int$, $\forall a.Erk~a \arrow a \arrow Int$ and
$\forall a. Erk~a \arrow a \arrow a$
(respectively the solutions from which they were derived)
do satisfy the above criteria. 
Hence, we cannot verify that they are not principal.

Instead, of checking for principality we simply compute all possible types
and check if one of these types is principal.
Our method is as follows.
Let ${\cal Q}.C_0 \wedge (D_1 \implies C_1) \wedge ... \wedge (D_n \implies C_n)$
be a normalized constraint generated out of expression $e$.
First, we check that ${\cal Q}.C_0 \wedge D_i \wedge C_i$ are satisfiable
for $i=1,..,n$ (we simply build the m.g.u.~under prefix ${\cal Q}$).
If not then either expression $e$ has a meaningless type annotation or
contains a type error. Note that the type error may be due to our limited
inference scheme. E.g., the constraints generated from Example~\ref{ex:size} 
via the rules Figure~\ref{f:type-inference} lead to an unsatisfiable
constraint although the program is well-typed.
Clearly, we need to report an error in such a situation and hope for further user input.
Otherwise, based on Lemma~\ref{le:finite-solutions} we compute
the sub-solutions ${\cal S}({\cal Q}.C_0 \wedge D_i \wedge C_i)$ 
for $i=1,..,n$ and compute via Lemma~\ref{le:enum-solutions}
all combinations which yield a solution.
Note that there can only be a finite number of solutions.
Hence, we can test whether any of these solutions is principal.

We can state the following result. 

\vspace{-1mm}
\begin{theorem}[Principal Types GRDTs]
We can infer a principal type for GRDTs if one exists and constraints
generated out of the expression are satisfiable.
\end{theorem}
\vspace{-1mm}

Note that based on our refined inference scheme in Section~\ref{sec:inf-criteria}
in combination with our method for building solutions
we find that function {\tt size} in Example~\ref{ex:size}
has types $\forall a.R~a \arrow Int$, $\forall a.R~a \arrow a$ but none of the two is principal.
Note that $R~Int \arrow Int$ is a meaningless type.
On the other hand, we find that {\tt eval} in Example~\ref{ex:eval} has the
principal type $\forall a. Exp~a \arrow a$.

\mysection{Conclusion and Related Work} \label{sec:conclusion}

To our knowledge,
there are only two previous works which study type inference for GRDTs.
The approach by Simonet and Pottier~\cite{simonet-pottier-hmg} uses the same abstraction
from program text to constraints in the first step of type inference.
They demand a sufficient number of type annotations such that solving
is tractable. In contrast, we could show that solving
is always tractable by reduction to unification under a mixed prefix.
We believe that our inference scheme will succeed for all programs which are successful
under their scheme.
They seem to imply that sufficient type annotations ensure that solving is tractable {\em and}
solving is successful. However, we can never rule out the event of a type error.

The goal of the work by Peyton-Jones, Washburn and Weirich~\cite{gad} is to make
type inference ``predictable''.\footnote{We would like to point out that no type inference system
is ever predictable due to (unavoidable) type errors in user programs.}
The gist of their work is to impose the condition
that if the type of the body of a pattern clauses changes due to a GRDT,
then the GRDT must be explicitly provided by the programmer.
Clearly, this condition is motivated by the fact that in a conservative inference scheme
we combine the results from the individual branches. Hence, we may fail
unless types are explicitly provided. However, they rule 
out Example~\ref{ex:grdt-ex2} which we have seen carries enough type information such that inference succeeds.

In this paper, we have introduced several 
improved inference methods for GRDTs
for guessing the types of GRDT programs (Section~\ref{sec:inference}), identifying missing information
based on the efficient inference criteria (Section~\ref{sec:inf-criteria}) and
building solutions via enumeration (Section~\ref{sec:enumerate}).
In combination, these methods allow us to infer the types of all examples in this paper.
Furthermore, we are the first to discuss extensively the issue of principal types.
We have presented novel methods to check if a type is not principal type
and to find a principal type if one exists (Section~\ref{sec:principal}).

In future work, we plan to investigate how our type debugging methods~\cite{interactive} developed for
Hindley/Milner typable programs can be adapted to the GRDT setting.

\newpage

\appendix

\mysection{Proofs}

\subsection{Proof of Theorem~\ref{th:trl-sound}}

Our assumptions are:
Let $\tenv$ an environment, $e$ an expression,
$F$ a constraint and $t$ a type such that $\tenv, e \trl (F \gd t)$.
Let $\psi={\it solve}(F)$.
Then $True, \psi(\tenv) \turns e : \psi(t)$.
\proofsin{
\begin{proof}[Sketch]
We can easily show that $F, \tenv, \turns e : t$ assuming we extend the sets of constraints allowed to appear
in judgments. Let $N$ be the normalization of $F$. Skolemization is a satisfiability preserving transformation.
Hence, if $\models \psi(N)$ then $\models \psi(F)$. We can easily 
verify that judgments are closed
under substitutions. Hence, we find that $True, \psi(\tenv) \turns e : \psi(t)$.
\end{proof}
}

\mysubsection{Proof of Lemma~\ref{le:suff-known-lemma}}

Our assumptions are:
Let $e$ be an expression containing no annotations and no nested case expressions.
Let ${\cal Q}.C_0 \wedge (D_1 \implies C_1) \wedge ... \wedge (D_n \implies C_n) \wedge K$
be the (normalized) constraint generated and $t$ the type of $e$.
Let $U$ be a (simple) user-provided constraint where $K_U = \wedge_{a\in\tv(U)} known(a)$.
If ${\cal K}(U \wedge K_U \wedge {\cal Q}.C_0 \wedge (D_1 \implies C_1) \wedge ... \wedge (D_n \implies C_n) \wedge K)=
      {\cal K}(U \wedge K_U \wedge {\cal Q}.C_0 \wedge (D_1 \implies C_1) \wedge ... \wedge (D_n \implies C_n))$
and $U\wedge {\cal Q}.C_0 \wedge (D_1 \implies C_1) \wedge ... \wedge (D_n \implies C_n)$ is satisfiable,
then $U\wedge {\cal Q}.C_0 \wedge (D_1 \implies C_1) \wedge ... \wedge (D_n \implies C_n)$
has a solution.

\proofsin{
\begin{proof}[Sketch]
We first consider the case that $U$ is $True$. Immediately, we find that
branches must agree. Otherwise, 
$\bigwedge_k (a_i =t_{i_k})$ is equivalent to $False$.
 Hence, $F_2 \models known(a_i)$. However, by assumption we have that 
${\cal K}(F_1)={\cal K}(F_2)$ and clearly $F_1 \not \models known(a_i)$ (not that $K_U$ is $True$ as well).
Hence, branches must agree.
Hence,  $\bigwedge_k (a_i =t_{i_k})$ is satisfiable.
We know that the constraint problem is satisfiable.
Hence, our efficient inference succeeds and generates a solution.

Assume $U$ is non-trivial. We assume the user-provided information is given by some type $t_0$.
W.l.o.g., we compare $F_1\equiv t=t_0 \wedge known(t_0) \wedge (D_i \implies C_i)$
against $F_2 \equiv t=t_0 \wedge (D_i \implies C_i) \wedge K$
where $K\equiv (known(a_i) \vee \bigwedge_k (a_i =t_{i_k})$. We ignore the prefix {\cal Q}.
Note that variables in $t_0$ are universally quantified.
We distinguish among the following two cases.

Case: Branches disagree, i.e.~type $a_i$ changes. Hence, $\bigwedge_k (a_i =t_{i_k})$ is equivalent to $False$.
 Hence, $F_2 \models known(a_i)$. By assumption
${\cal K}(F_1)={\cal K}(F_2)$, hence, $a_i$ is defined in $t=t_0 \wedge known(t_0)$, i.e. $a_i \in \tv(t_0)$.
By assumption the constraint generated is satisfiable. Hence,  
a solution $\phi$ of $t=t_0 \wedge (D_i \implies C_i)$ exists. We build the m.g.u.~$\psi$ of $t=t_0$. Hence, $\psi \leq \phi$,
i.e.~$\psi$ is more general than $\phi$.
In particular, we have that $\psi(a_i) \leq \phi(a_i)$ (2). We consider the efficient inference problem
$t=t_0 \wedge E_i$. Note that $\phi$ is a solution. Because of (2) we also have that $\psi$ is a solution (for all $a_i$'s
which change their types in different branches). Hence, efficient inference succeeds.

Case: Branches agree. Hence,  $\bigwedge_k (a_i =t_{i_k})$ is satisfiable.
Same reasoning as before shows that efficient inference succeeds.
\end{proof}
}

\mysubsection{Proof of Lemma~\ref{le:enum-solutions}}

Our assumptions are:
Let ${\cal Q}.C_0 \wedge (D_1 \implies C_1) \wedge ... \wedge (D_n \implies C_n)$ be
such that $\psi$ is a solution and
${\cal Q}.C_0 \wedge D_i \wedge C_i$ is satisfiable for $i=1,..,n$.
Then, there exist $S_i \in {\cal S}({\cal Q}.C_0 \wedge D_i \wedge C_i)$ for $i=1,..,n$
such that $E_{\psi}$ and $\bigwedge_{i=1,...,n} S_i$ are equivalent w.r.t. $\tv({\cal Q}.F)$.

\proofsin{
\begin{proof}[Sketch]
We abbreviate $E_{\psi}$ by $S$.
We have that $S \implies {\cal Q}.F$ iff ${\cal Q}. S\implies C_0 \wedge (S\wedge D_1 \implies C_1) \wedge ...
\wedge (S \wedge D_n \implies C_n)$ (assuming bound variables have been renamed). Let $V=\tv({\cal Q}.F)$.
Clearly, we have that $S$ when projected onto $V$ is contained in $\bigcup_{i=1,...,n} {\cal S}({\cal Q}.C_0\wedge D_i \wedge C_i)$.
\end{proof}
}

\mysubsection{Proof of Lemma~\ref{le:necessasry-criteria}}

Our assumptions are:
Let $\psi$ be a principal solution of $F$.
Let $F'$ be the skolemized version of $F$ of the form $C_0,(D_1 \implies C_1),...,(D_n \implies C_n)$.
Then, 
$(E_{\psi},C_0,\bigwedge_{i=1,...,n} D_i) \leftrightarrow 
      (C_0, \bigwedge_{i=1,...,n} (C_i, D_i))$ where 
$E_{\phi}= \{ a=\phi(a) \mid a\in {\it domain}(\psi) \}$.

\proofsin{
\begin{proof}[Sketch]
Note that $\psi$ is a solution of $F$ iff $E_{\psi} \implies F$.
Note that skolemization is a satisfiability maintaining transformation.
Hence, we can assume that $E_{\psi} \implies F'$ (for convenience we keep
the implicit universal quantifier).
In the following, we use $S$ as a short-hand for $E_{\psi}$.
We have that $C_0, \bigwedge_{i=1,...,n} C_i$ is a solution.
$\psi$ is principal, hence, $C_0,\bigwedge_{i=1,...,n} C_i \implies S$  (1).
From (1), we obtain that $C_0, \bigwedge_{i=1,...,n} (D_i, C_i) 
   \implies S, C_0, \bigwedge_{i=1,...,n} D_i$ (2).
$\psi$ is a solution, hence, $S, C_0 \implies C_0, \bigwedge_{i=1,...,n} (D_i \implies C_i)$.
We conclude that $S, C_0, \bigwedge_{i=1,...,n} D_i \implies \bigwedge_{i=1,...,n} C_i$ (3).
From (2) and (3), we obtain that $(S, C_0, \bigwedge_{i=1,...,n} D_i) \leftrightarrow 
      (C_0, \bigwedge_{i=1,...,n} (C_i, D_i))$.
\end{proof}
}

\end{document}